\documentclass{elsarticle}
\usepackage[utf8]{inputenc}
\usepackage{graphicx}
\usepackage{amsfonts}
\usepackage{amssymb}
\usepackage{mathtools}
\usepackage{subcaption}
\usepackage{xcolor}
\usepackage{url}
\usepackage{soul}
\usepackage{comment}
\usepackage[normalem]{ulem}
\usepackage{multirow}
\usepackage{tikz}
\usetikzlibrary{arrows, positioning, shapes}
\usetikzlibrary{fit}
\usetikzlibrary{backgrounds}
\journal{ }
\usepackage{algorithm}
\usepackage{algpseudocode}

\begin{document}

\title{Equivariant graph neural network surrogates for predicting the properties of relaxed atomic configurations}

\author[umap]{Jamie Holber\corref{cor1}}
\ead{holber@umich.edu}

\author[auaero]{Siddhartha Srivastava}
\author[usc]{Krishna Garikipati}

\cortext[cor1]{Corresponding author}

\address[umap]{Applied Physics, University of Michigan, Ann Arbor, USA}
\address[auaero]{Department of Aerospace Engineering, Auburn University, Auburn, USA}
\address[usc]{Department of Aerospace and Mechanical Engineering, University of Southern California, Los Angeles, USA}

\begin{abstract}

{Density {f}unctional {t}heory (DFT) is the workhorse method for electronic structure-based computations of material properties. However,} {due to the computational expense of DFT, many methods are being explored to utilize fewer DFT computations without sacrificing accuracy.}   Equivariant graph neural networks (EGNNs) represent {one such approach which} provide a flexible framework for predicting material properties that inherently respects the symmetry of the system and is not restricted to a particular lattice.  In this work, we present {an EGNN-based}  mathematical framework to predict {key structure–property relationships relevant to solid-state battery cathode materials} 
including formation energy, strain, and atomic displacements. We apply this framework to the system Li$_x$CoO$_2${, a layered oxide cathode exhibiting lithium intercalation and diffusion,} across various compositions {(x) and configurations} of Li. Our results demonstrate that the EGNNs can accurately predict quantities outside the training set including the largest atomic displacements, the strain tensor and {strain} energy {density}, and the formation energy, providing greater insight to the{ interplay between electrochemistry and lattice distortions}  without the need for more DFT calculations.

\end{abstract}
\maketitle

\section{Introduction}

The efficient and accurate prediction of material properties is a central goal in computational materials science. For many material systems, including crystalline solids, density functional theory (DFT) has been the principal method for calculating properties such as formation energy, band gaps, and elastic properties. Schleder et al. provide a recent overview of the role of DFT in computational materials science, including its integration with complementary modeling approaches \cite{schleder2019dft}. Although DFT can achieve high accuracy, it is computationally expensive, particularly for large or complex systems. As a result, machine learning (ML) methods have emerged as powerful tools for extracting more information from limited DFT datasets. For example, Nyshadham et al. demonstrated that ML models trained on DFT data can serve as surrogate models to rapidly predict material properties for structures outside the training set \cite{nyshadham2019machine}. Popular surrogate approaches include regression methods \cite{ulissi2017address}, cluster expansions \cite{Thomas2013Anharmonic,Puchala2013ZrO}, and neural networks. In particular, equivariant graph neural networks (EGNNs) have become increasingly popular for modeling atomistic systems \cite{shi2024review}.

Reiser et al. showed that EGNNs can achieve high accuracy in predicting material properties while respecting the underlying symmetries of atomic configurations \cite{reiser2022graph}. In these models, atomic structures are represented as graphs, where nodes correspond to atoms and edges represent interatomic interactions \cite{shi2024review}. Structural and chemical information is encoded through node, edge, and global features. Node features may include elemental properties such as group number, period number, and electronegativity \cite{PhysRevLett.120.145301}, while edge features can incorporate geometric descriptors such as interatomic distances and angular relationships, as demonstrated by Batzner et al. Global features may represent system-level quantities such as strain, stress, and deformation, which can also be predicted within this framework \cite{maurizi2022predicting}.

Many EGNN-based models have been developed for general materials screening, such as those presented by Ojih et al. \cite{ojih2024graph} and Boonpali et al. \cite{boonpalit2023data}. These models are typically trained on large and diverse datasets and are used to predict properties of novel materials. A non-exhaustive list of prominent interatomic potentials includes ORB \cite{rhodes2025orbv3atomisticsimulationscale,neumann2024orbfastscalableneural}, CHGNet \cite{deng2023chgnet}, NequIP \cite{batzner20223}, and MACE \cite{Batatia2022mace}. The crystal graph convolutional neural network (CGCNN) is a widely used model for crystalline materials, achieving a mean absolute error (MAE) of 0.039 eV in predicting formation energies \cite{shi2024review,PhysRevLett.120.145301}. While such models offer strong accuracy and generalizability across diverse materials, they may be less suitable for modeling a specific material system, where meV or sub-meV accuracy is required.

In our previous work, we developed a scale-bridging workflow to study phase-field dynamics in lithium cobalt oxide (LCO), where the free energy density was represented by a neural network trained on data obtained from Monte Carlo (MC) simulations \cite{shojaei2024bridging}. In that framework, DFT calculations were used to parameterize a cluster expansion of the formation energy, which subsequently informed the MC simulations. Cluster expansions, such as those implemented in CASM \cite{Thomas2013Anharmonic,Puchala2013ZrO,VanderVen2018StatMech,casm}, have proven to be powerful tools for modeling the relationship between configurational states and crystal structure, even in complex systems \cite{wu2016cluster,kadkhodaei2021cluster}. In our study of LCO, the cluster expansion achieved an accuracy on the order of a few meV. However, cluster expansions have several limitations. They are typically restricted to crystalline solids and often rely on unrelaxed atomic configurations. This can introduce inconsistencies when trained on unrelaxed structures but fitted to DFT energies of relaxed configurations. Furthermore, cluster expansions do not naturally account for defects or amorphous structures.

In this work, we develop an EGNN framework that, rather than targeting broad generalization across materials, is specialized to model different configurations of a single material system with higher accuracy. Witman et al. \cite{witman2024phase} demonstrated that graph neural networks can predict formation energies for both relaxed and unrelaxed structures with errors on the order of a few meV using CGCNN. Building on this idea, we construct an EGNN that predicts not only the formation energy, but also the effective strain tensor (yielding relaxed lattice parameters) and the relaxed atomic positions. This approach enables approximation of the DFT relaxation process itself, in addition to predicting formation and strain energies. These quantities are critical inputs for higher-scale models, including phase-field simulations of evolving microstructures with elastic inhomogeneity \cite{hu2001phase,Miehe2010} and multiphase systems \cite{steinbach2006multi}.

We first present the computational framework in Section \ref{sec:methods}, including data generation via DFT and an overview of cluster expansions, followed by a description of the EGNN architecture. In Section \ref{sec:results}, we present the results of the DFT calculations and the performance of the trained EGNN.

\section{Methods} \label{sec:methods}

\subsection{Density Functional Theory}

Density functional theory is a quantum mechanical modeling framework for investigating the electronic structure of many-body systems. By approximating the many-electron wavefunction using functionals of the electron density, DFT enables the computation of key material properties, such as formation energy, charge density distribution, and electronic band structure. One of the fundamental objectives in DFT calculations is to determine the ground-state configuration of a system—that is, the {spatial distribution of atoms and electron density that minimizes the total energy.}

\begin{figure}
    \centering
    \includegraphics[width=0.5\linewidth]{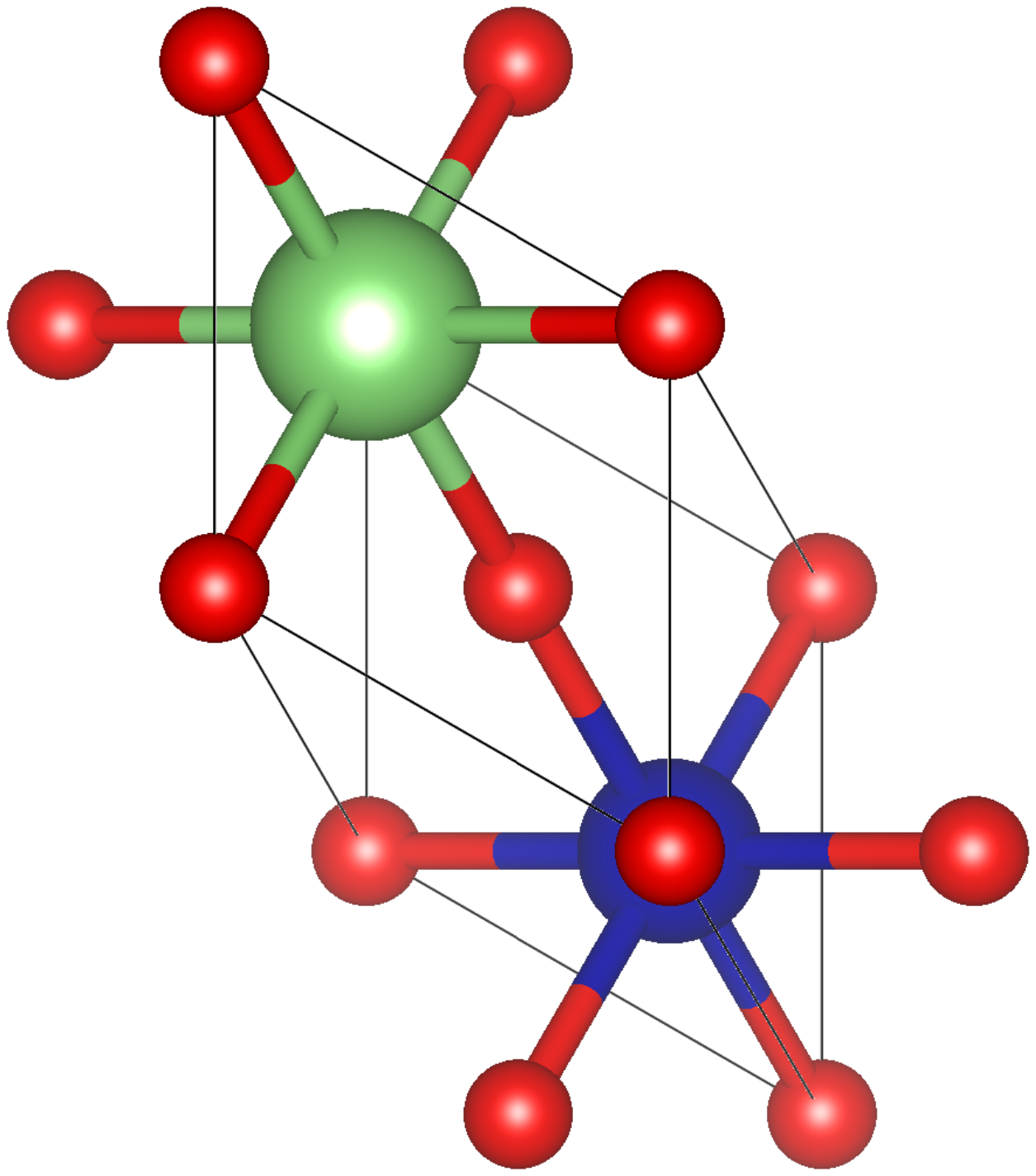}
    \caption{The primitive cell of LCO. Red is oxygen, green is Lithium, and blue is Cobalt.}
    \label{fig:primitive}
\end{figure}

The relaxed DFT configurations provide the equilibrium lattice vectors and atomic coordinates, from which a primitive cell is constructed by identifying the symmetrically distinct crystallographic sites corresponding to the most-probable atomic positions. {From this primitive cell, a set of supercells is formed, where each supercell} {is a  combination of  an integer number of primitive cells with varying occupations of the lattice sites. Thus each supercell can have a different composition and tiling of Li.} For each supercell, DFT computations typically begin with an initial, “unrelaxed” atomic configuration.  {DFT relaxation computations adjust the positions of atoms and the lattice parameters to determine the structure with the lowest total energy by } iterative electronic and ionic relaxations, updating the electron density and atomic positions at each step until the energy meets a convergence criterion. The total formation energy is computed for each configuration, providing insight to the thermodynamic stability of different atomic arrangements. {The total energy for each configuration, $E_{tot}$, computed by DFT is then used to calculate the formation energy as:}

\begin{equation}
    E_f = \frac{E_{tot}}{M} - E_{adjust}
    \label{eq:form_energy}
\end{equation}
where $M$ is the size of the configuration in terms of number of primitive cells and $E_\text{adjust}$ is
\begin{equation}
    E_\text{adjust} = (1-x)*E_{\text{CoO}_2} + x*E_{\text{LiCoO}_2}
\end{equation}
where $x$ is the composition of Li. $E_{\text{CoO}_2}$ is the energy of the primitive cell with no lithium and $E_{\text{LiCoO}_2}$ is the energy of the primitive cell with lithium.  This ensures that the formation energy is 0 at $x=0$ and $x=1$. {The primitive cell is shown in Figure \ref{fig:primitive}}

In this work, we perform DFT computations with an on-site Hubbard $U$ correction (DFT$+U$) and van der Waals (vdW) functionals to account for strong electron correlations and long-range dispersion forces, respectively. These corrections are particularly important for accurately modeling transition metal oxides, such as LCO, where localized $d$-electron interactions and interlayer vdW interactions play significant roles in determining structural and electronic properties.

DFT computations, while generally accurate, are also expensive, so alternate, accurate methods are explored to predict formation energies for different configurations

\subsection{Cluster Expansions}


In previous work we have have employed cluster expansions, {trained to DFT data,} to model the {formation energy's dependence} on atomic configuration \cite{shojaei2024bridging}. {To train a cluster expansion, we begin by defining an occupancy array $\boldsymbol{\sigma} = \{\sigma_1,\sigma_2,...,\sigma_n\}$ for each structure in our dataset, where each $\sigma_i$ is the occupancy variable corresponding to a lattice site $i$. Here, $\sigma_i =1$ if  site $i$ is occupied and $\sigma_i = 0$ otherwise. Given the lattice, we then define a set of clusters $\mathcal{C}$ on the Li sub-lattice, where each cluster $(\alpha \in \mathcal{C} )$ is a subset of the set of all sites of a structure, defined as $\alpha = \{i,j,...k\}$ and we can define $\phi_\alpha = \sigma_i\sigma_j...\sigma_k$, where $\phi_\alpha = 0$ if any site in the cluster is unoccupied. While there is a commonality of cluster types across structures, not every cluster appears in each structure due to differences in the cell size and shape.}     We define a maximum number of sites per cluster and a {maximum distance between sites, reached by traversing an integer number of bonds along a path in the lattice}. Then a cluster expansion is a {polynomial} defined by a linear combination of the {monomials} $\phi_\alpha$. 
\begin{equation}
    E_f(\boldsymbol{ \sigma}) = V_0 + \sum_{\alpha \in \mathcal{C}} V_\alpha \phi_\alpha(\boldsymbol{\sigma})
    \label{eq:clex}
\end{equation}

The coefficients $V_0$ and all $V_\alpha $ are optimized using a sparse regression technique {that weights a configuration, $i$ based upon the energy difference $\Delta E_c$ {(eV/unit cell)} relative to the configuration on the convex hull with the same Li composition $x$ using} {the following weights}  
\begin{equation}
    w_\sigma = 15 \exp (-\frac{\Delta E_\sigma^{\text{hull}}}{0.005})+0.5
\end{equation}
where $\Delta E^\text{hull}_{\boldsymbol{\sigma}}$ is the energy difference to the convex hull for that configuration. {This ensures that the predicted energy $E_f(\boldsymbol{ \sigma})$ in Eq (\ref{eq:clex}) has higher accuracy for the lowest energy configurations.} 

Cluster expansions are well proven for predicting formation energies. However, {traditional cluster expansions} rely on the structure being crystalline and do not allow for flexibility in the configuration{, such as out-of-equilibrium  positions of atoms due to structural relaxations induced by vacancies}.  {This motivates our interest in GNN approaches.}

\subsection{Graph Neural Networks for Atomic Systems}

In this work, we develop an equivariant graph neural network (EGNN) that can learn formation energies, {as well as lattice strain and atomic displacement from DFT data, thus predicting relaxed geometries from the unrelaxed structure.}  {We begin with an outline of the EGNN and develop the mathematical formulation in the following subsections.} 

\begin{figure}
    \centering
    \includegraphics[width=\linewidth]{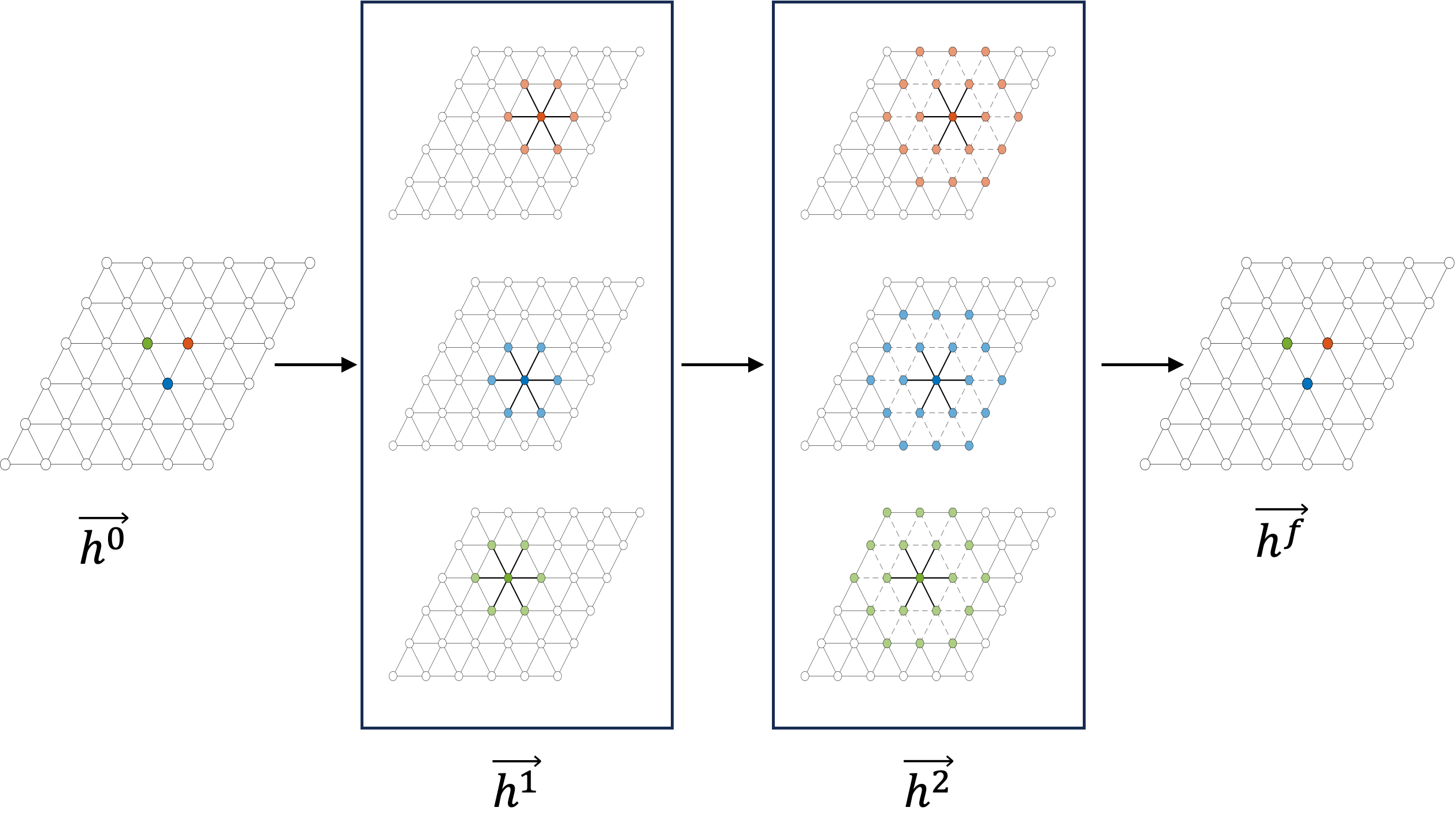}
    \caption{Illustration of message passing in a graph neural network (GNN) on a triangular lattice. Solid black edges represent direct connections between neighboring sites, while colored nodes indicate the receptive field of the central node at different layers. As the number of layers increases, the receptive field expands to include more distant neighbors. }
    \label{fig:GNN}
\end{figure}

The graph is defined such that the edge features incorporate both interatomic distances and angles, and the {vertex} features incorporate the type of atom.  While the graph structure remains the same in each layer, the {vertex} features are updated using convolutional mechanisms, which allows the EGNN to learn the atomic features based on the local environment using message passing. In each message-passing step, messages are computed along the edges of the graph by applying a learned function, which is dependent on a set of weights, the edge features, and the {vertex} features on either end of the edges. Then at each {vertex}, all the messages for edges corresponding to that {vertex} are aggregated, typically using summation or averaging, and used to update the {vertex}'s feature. This process allows the network to build a representation of a {vertex}'s local environment. {This architecture generalizes the convolutional neural networks (CNNs) from their application on structured pixelated data to unstructured graph data.} Unlike multilayer perceptrons, convolutional neural networks (CNNs) apply the same set of weights and biases across a layer, so the message passing functions will be the same for each {vertex}. By stacking multiple layers, the network can propagate information across the structure, enabling longer-range interactions to influence the local predictions. This allows our model to make precise predictions for atomic relaxations and formation energies. Figure \ref{fig:GNN} shows an example of a graph neural network with 2 hidden layers, highlighting a few vertices. {It can be observed that by adding multiple layers in the EGNN, the messgage passing carries aggregated message beyond the first neighbors, thus introducing a systematic long-range effect in the receptive field.}

In our EGNN implementation, {the convolutional operation is E(3)-Equivariant, {such that the graph is invariant under rotations, translations, and reflections, The predicted energy and strain tensor are invariant, while the predicted displacements in positions  {satisfy an equivariance introduced in the final output layer.} The E in E(3) refers to Euclidean},   noting that the predictions {are invariant to the} arbitrary choice of coordinate system.} This is especially important given that different DFT configurations could have different coordinate systems. {Enforcing these symmetries allows for better learning of representations. }

\subsubsection{Graph Construction} \label{Graph construction}

Each crystal structure is represented by a graph $G= ( \mathcal{V}, \mathcal{E} )$ where $\mathcal{V}$ corresponds to the atoms and $\mathcal{E}$ corresponds to the {inter-atomic attributes}. Each {vertex} has a feature, which initially corresponds to the atomic type. In this work we set the {vertex} features{, for the input layer L=0} as $\mathbf{h}_i^0=[1]$ for Li, $\mathbf{h}_i^0=[2]$ for Co, and $\mathbf{h}_i^0=[3]$ for O. Additionally, each edge has the set of  $e_{ij}$ features containing the interatomic distance and the angular differences between neighboring pairs $e_{ik}$ and $e_{ij}$. {The edges are directional due to  differences in the angular component.}  To calculate the interatomic distances we consider the relationship between two {vertices}:

\begin{equation}
\mathbf{r}_{ij} = \mathbf{r}_j - \mathbf{r}_i
\end{equation}

However, in DFT it is assumed that the supercells are infinitely repeating under periodic boundary conditions (PBC). Therefore, the atom at position $\mathbf{r}_i$ not only interacts with another atom $j$ at position $\mathbf{r}_j$, but also interacts with all the periodic images of atom $j$. To compute interatomic interactions correctly it is standard to use the shortest distance between atom $i$ and any periodic image of $j$--known as the minimum image convention. This periodic image is located in the original {supercell}  or in an adjacent {supercell}. The transformation to the periodic images is given by $\mathbf{T}_s$:
\begin{equation}
    \mathbf{T}_s = \mathbf{s} \cdot\mathbf{L}, 
\end{equation}
where $\mathbf{L}$ is the matrix representation of the lattice vector parameters given by
\begin{equation}
    \mathbf{L} = [\mathbf{l}_x, \mathbf{l}_y, \mathbf{l}_z]
\end{equation}
where $\mathbf{l}_x, \mathbf{l}_y, \mathbf{l}_z$ are the lattice vectors defining the unit cell and {over which the minimum image convention holds.}
\begin{equation}
\mathbf{s} = [s_x,s_y,s_z],\quad s_x,s_y,s_z \in \{-1,0,1\}
\end{equation} 

This yields all 27 relevant unit cells, including the original. {These} periodic images in the supercells adjacent to the original {can be represented as} 
\begin{equation}
    \mathbf{r}_j^{(s)} = \mathbf{r}_j + \mathbf{T}_s.
\end{equation}

Then the smallest distance can then be calculated as: 
\begin{equation}
 d_{ij} = \min (||\mathbf{r}_j^{(s)} - \mathbf{r}_i||,\forall \mathbf{s} = [s_x,s_y,s_z],\quad s_x,s_y,s_z \in \{-1,0,1\}
\end{equation}
We only define an edge if $d_{ij} < d_\text{min}$, a suitable threshold tolerance. { The complete set of edges $(E)$ is determined by considering all distinct pairs of vertices and selecting those that satisfy the threshold condition.
\begin{equation}
    \mathcal{E} = \{ e_{ij} | i,j \in V, i\neq j, d_{ij} < d_\text{min} \}
\end{equation}
}

{This edge $e_{ij}$ consists of the minimum image distance $d_{ij}$ and an angular edge attribute $\alpha_{ij}$. This angular edge attribute is determined by first defining the unit edge vector:}

\begin{equation}
\hat{\mathbf{r}}_{ij} = \frac{\mathbf{r}_{ij}}{\Vert\mathbf{r}_{ij}\Vert}
\end{equation}
{Then for each  atom $k$ neighboring atom $i$, but $k\neq j$ for $j$ also a neighbor of $i$,} we compute the bond angle:
\begin{equation}
\cos\theta_{ijk} = \hat{\mathbf{r}}_{ij} \cdot \hat{\mathbf{r}}_{ik}
\end{equation}
\begin{equation}
\theta_{ijk} = \cos^{-1} (\cos\theta_{ijk})
\end{equation}
We then define an aggregated angular descriptor for each edge: 
\begin{equation}
\beta_{ij} = \frac{1}{n-1} \sum_{k \neq j} \theta_{ijk}
\end{equation}
where $n$ is the number of neighbors of vertex $i$. Thus, the final edge attributes are represented as 
\begin{equation}
\mathbf{e}_{ij} = [d_{ij}, \beta_{ij}]
\end{equation}
Since the  {features $\mathbf{h}_i$, $\boldsymbol{e}_{ij}$ are only} dependent on the distance and angles between {vertices}, the {graph representation} is {invariant}{ under the symmetry operations of rotation, translation and reflection}. {While the predicted energy and strain tensor are also invariant, the predicted change in positions has equivariance imposed in the output layer.} 

\subsubsection{EGNN Message Passing}
\label{sec:msgpass}

Once the graph structure is established, we apply message-passing to update the {vertex} representations. The message passing function $f$ at layer $L$ for each edge is defined as:

\begin{equation}
m^L_{ij} = f(\mathbf{h}^L_i,\mathbf{h}^L_j,\mathbf{e}_{ij})
\label{eq: message mlp}
\end{equation}
where $f$ is a learnable {scalar} function (MLP), shown in Table \ref{tab:MLP} and represented in Figure \ref{fig:MLP_messagepassing}. The function $f$ does not change depending on the {vertex}, but will have the same weights for each {vertex} in a given layer. Next, each {vertex} aggregates messages from its neighbors {using an averaging operation}:

\begin{equation}
\mathbf{m}^L_i = \frac{1}{n} \sum_j {m}^L_{ij}
\end{equation}

Finally, the {vertex} features are updated using {the vertex update} {MLP} {representation} $g$, { the exact form of which is shown in Table \ref{tab:MLP} and illustrated in Figure \ref{fig:MLP_messagepassing} }

\begin{equation}
\mathbf{h}^{L+1}_i = g([\mathbf{h}_i^L, \mathbf{m}^L_i])
\label{eq : vertex MLP}
\end{equation}
 Note that for each layer the message $\mathbf{m}^L_i$ and {vertex} features $\mathbf{h}^L_i$ can each be a vector of multiple values. They therefore appear in boldface font.

{Given $L_h$ hidden layers, the final vertex features are represented by $\mathbf{h}_i^{L_h}$. Then we have a final output layer which predicts the formation energy, relaxed lattice strain, and atomic positions respectively: 
\begin{equation}
    \text{Output} = E_f^{\text{Pred}}( \mathbf{h}_i^{L_h}), \mathbf{E}^{\text{Pred}}(\mathbf{h}_i^{L_h}), P^{\text{Pred}}_{\delta}(\mathbf{h}_i^{L_h})
\end{equation}

where {the exact functions are defined in the next section}

\subsubsection{Output Parameters}

Our model is designed to simultaneously predict three key outputs from the initial unrelaxed structure: the formation energy, the relaxed lattice strain, and the atomic position displacements after relaxation.  The first output we consider is the formation energy ($E_f$) of the final relaxed structure, which can be computed from the DFT results as shown in Equation (\ref{eq:form_energy}) and is defined as the ground truth for training: $E_f^{\text{True}}$. To determine the predicted formation energy, we begin by {determining the mean of} all the local {vertex} features ($\mathbf{h}^{L_h}$) to get a global feature vector {for all sites $N$}. 

\begin{equation}
    \mathbf{h}_{\text{out}} = \frac{1}{N}\sum \mathbf{h}_i^{L_h}
    \label{eq:global features}
\end{equation}
Then we pass the output through a {formation energy} MLP {($f_{E_f}$)}, {described in Table \ref{tab:MLP} and consisting of the first two layers of Figure \ref{fig:MLP_messagepassing}}, which has an output of size 1, and dividing by the size of the configuration \sout{($V$)} {($M$)}: 
\begin{equation}
    E_f^{\text{Pred}} = \frac{f_{E_f}( \mathbf{h}_{\text{out}})}{M}
    \label{eq:efoutput}
\end{equation}
{Note that $M$ is the integer multiple of the primitive cell, which can include vacancies, and differs from $N$, which is the number of occupied lattice sites.}
The formation energy loss is then calculated as:
\begin{equation}
    \mathcal{L}_{E_f} = \mathrm{MSE}(E_f^{\text{Pred}},E_f^{\text{True}})
\end{equation}
In addition to {predicting} formation energy, {we also seek to predict the change in the lattice parameters under DFT relaxation, as represented by a deformation gradient}. 
Given the initial {matrix representation of the} lattice vectors $\mathbf{L^0}${$=[\mathbf{l^0_x},\mathbf{l^0_y},\mathbf{l^0_z}] $} and the {matrix representation of the} deformed {(ie DFT-relaxed)} lattice vectors {$\mathbf{L^d} =[\mathbf{l^d_x},\mathbf{l^d_y},\mathbf{l^d_z}] $}, then the deformation gradient {is the linear transformation satisfying}:
\begin{equation}
    \mathbf{L^d} = \mathbf{F}\mathbf{L^0}
\end{equation}
Therefore, the deformation gradient can be calculated as:
\begin{equation}
    \mathbf{F} = \mathbf{L^d}\mathbf{(L^{0})^{-1}}
\end{equation}
The deformation gradient tensor $\mathbf{F}$ is a fundamental descriptor {of kinematics} in continuum mechanics that is not{ frame invariant, and is therefore not an an objective representation of strain}. Therefore, we instead use the Green-Lagrange strain: 
\begin{equation}
    \mathbf{E} = \frac{1}{2}(\mathbf{F}^T\mathbf{F}-\mathbf{I})
    \label{eq:green-lagrange}
\end{equation}
Using the St. Venant-Kirchhoff model of nonlinear elasticity, the strain energy density can then be written in terms of the elasticity tensor $\mathbb{C}$, {and using coordinate notation}:
\begin{equation}
    U (\mathbf{E}) = \frac{1}{2} \mathbb{C}_{IJKL} {E}_{IJ}{E}_{KL}
    \label{eq:strain energy calc}
\end{equation}
The strain energy per primitive cell is then:
\begin{equation}
    S = U/M
\end{equation}
 To determine the loss function for strain energy, we define a loss based on the difference between the predicted and true strain tensor. The true strain tensor $\mathbf{E}^{\text{True}}$, {which is symmetric}, is calculated using Equation (\ref{eq:green-lagrange}) and the lattice parameter {vectors obtained} from DFT. The  strain tensor is  {predicted}  using the global features tensor {as shown in} Equation (\ref{eq:global features}) {and running it} through {the lattice} MLP{, as shown in Table \ref{tab:MLP}, and illustrated in Figure \ref{fig:MLP_messagepassing}} to yield a tensor {representing the predicted Green-Lagrange strain.} 
\begin{equation}
    \mathbf{E}^{\text{Pred}} = f_{l}(\mathbf{h}_{\text{out}}) \in \mathbb{R}^{3\times 3}
    \label{eq:strainoutput}
\end{equation}
{Then the difference between the true (from Equation \ref{eq:efoutput})  and predicted strain is calculated:
\begin{equation}
    \mathbf{E}^\delta = \mathbf{E}^{\text{True}} -  \mathbf{E}^{\text{Pred}}
\end{equation}
In order to penalize the strain discrepancy $\mathbf{E}^\delta$, we use the induced strain energy density calculated using Equation \ref{eq:strain energy calc} per primitive cell:
\begin{equation}
    S = \frac{U(\mathbf{E}^\delta)}{M}
\end{equation}
and the corresponding loss function is:
\begin{equation}
    \mathcal{L}_S = \mathrm{MSE}(S,0)
\end{equation}
This encourages the model to minimize the predicted strain discrepancy, emphasizing the terms with high stiffness while avoiding fortuitous cancellations which could occur if the discrepancy in strain energy was instead used.

{While the strain tensor captures the change in the lattice parameters as the   structural deformation homogenized over the entire lattice, individual, internal atoms can undergo additional motions that have no effect on the overall lattice geometry.}
These {internal} changes in the structure provide insight about interatomic distances between atoms and the stability of a certain structure. These atomic displacements are quantified as the difference between the set of Cartesian coordinates in the original structure $\mathbf{r}^{\text{unrelaxed}}$ and the DFT predicted final set of Cartesian coordinates $\mathbf{r}^{\text{relaxed}}$.:
\begin{equation}
    \mathbf{r}^{\delta,\text{True}}=\mathbf{r}^{\text{unrelaxed}}-\mathbf{r}^{\text{relaxed}}
\end{equation}
We wish to use the EGNN to predict this change in position $\mathbf{r}^{\delta,\text{True}}_i$, {accounting for their changes} under rotations. To ensure that the network {possesses proper} E(3) symmetry, we {use} the relative positions of neighboring atoms as given by the bond vector, $\mathbf{r}^{\text{unrelaxed}}_i-\mathbf{r}^{\text{unrelaxed}}_j$. The predicted change in position is then given by:

\begin{equation}
   \mathbf{r}^{\delta,\text{Pred}}_i = \sum_{j \neq i} (\mathbf{r}^{\text{unrelaxed}}_i - \mathbf{r}^{\text{unrelaxed}}_j) f_r(\mathbf{h}_i^{L_h},\mathbf{h}_j^{L_h},\mathbf{e}_{ij}) 
    \label{eq:outputpos}
\end{equation}
{Here, $f_{r}$ is {the coordinate} MLP, described in Table \ref{tab:MLP} and illustrated in Figure \ref{fig:MLP_messagepassing}, which outputs the atomic displacement along each edge. This formulation ensures equivariance such that if the entire structure is rotated by some $\mathbf{R}$ then both the predicted change in coordinates $\mathbf{r}^{\delta,\text{Pred}}_i$ and the true change in coordinates $\mathbf{r}^{\delta,\text{True}}_i$ will {inherit this rotation} without $f_r$ needing to change. {The output of $f_r$ is a scalar so that  $(\mathbf{r}^{\text{unrelaxed}}_i - \mathbf{r}^{\text{unrelaxed}}_j)f(r) $ yields a vector of dimension 3.}

 The positions loss is then calculated as:
\begin{equation}
    \mathcal{L}_r = \frac{1}{N}\sum_{i\in N}\mathrm{MSE}( \mathbf{r}^{\delta,\text{Pred}}_i, \mathbf{r}^{\delta,\text{True}}_i)
\end{equation}
The total loss then combines the three losses, where $a,b,\text{ and } c$ are {positive} constants. 
\begin{equation}
    \mathcal{L} = a*\mathcal{L}_{E_f} + b*\mathcal{L}_S + c*\mathcal{L}_r
\end{equation}
The representation of the entire EGNN is shown in Figure \ref{fig:EGNN_schematic}. {For each message passing layer, the message passing function $f$ and the vertex update function $g$ are parameterized by the same weights that are identical for nodes and edges across the graph for that layer.}

\begin{table}[h!]
\centering
\begin{tabular}{|c|c|c|c|c|}
\hline
\textbf{MLP} & \textbf{Equation Ref} & \textbf{Layer} & \textbf{Input Dimension} & \textbf{Output Dimension} \\
\hline

\multirow{3}{*}{Message {Passing} $f$} 
 & \multirow{3}{*}{\parbox{3cm}{\centering $ f(\mathbf{h}^L_i,\mathbf{h}^L_j,\mathbf{e}_{ij})$ \\ in Eq.~\ref{eq: message mlp} }} 
 & Linear & $2 \cdot d_\text{vertex} + d_\text{edge}$ &  $d_\text{hidden}$ \\
 & & ReLU   &  $d_\text{hidden}$ &  $d_\text{hidden}$ \\
 & & Linear &  $d_\text{hidden}$ &  $d_\text{hidden}$ \\
\hline

\multirow{4}{*}{Vertex Update $g$} 
 & \multirow{4}{*}{\parbox{3cm}{\centering $ g( [\mathbf{h}^L_i,\mathbf{m}^L_i)]$ \\ in Eq.~\ref{eq : vertex MLP}}} 
 & Linear & $d_\text{vertex} + d_\text{hidden}$ &  $d_\text{hidden}$ \\
 & & ReLU  &  $d_\text{hidden}$ &  $d_\text{hidden}$ \\
 & & Linear &  $d_\text{hidden}$ &  $d_\text{hidden}$ \\
 & & ReLU &  $d_\text{hidden}$ &  $d_\text{hidden}$ \\
\hline

Formation Energy $f_{E_f}$ 
 & \parbox{3cm}{\centering $f_{E_f}(\mathbf{h}_\text{out})$ \\ in Eq.~\ref{eq:efoutput}} 
 & Linear &  $d_\text{hidden}$ & 1 \\
\hline

\multirow{3}{*}{Lattice $f_{l}$} 
 & \multirow{3}{*}{\parbox{3cm}{\centering $f_{l}(\mathbf{h}_\text{out})$ \\ in Eq.~\ref{eq:strainoutput}}} 
 & Linear &  $d_\text{hidden}$ & $d_\text{lattice}$ \\
 & & ReLU   & $d_\text{lattice}$ & $d_\text{lattice}$ \\
 & & Linear & $d_\text{lattice}$ & $9$ (reshaped to $3 \times 3$) \\
\hline

\multirow{3}{*}{Coordinate $f_r$} 
 & \multirow{3}{*}{\parbox{3cm}{\centering $f_r(\mathbf{h}_i^{L_f},\mathbf{h}_j^{L_f},\mathbf{e}_{ij})$ \\ in Eq.~\ref{eq:outputpos}}} 
 & Linear &  $d_\text{hidden}$ & $d_\text{coords}$ \\
 & & ReLU   & $d_\text{coords}$ & $d_\text{coords}$ \\
 & & Linear & $d_\text{coords}$ & $1$ \\
\hline

\end{tabular}
\caption{{Summary of all MLP architectures used in the E-3 GNN. $d_\text{vertex}$ is the number of components in the feature vector for each vertex, $d_\text{edge}$ is the number of components in the feature vector for each edge,  $d_\text{hidden}$ is the number of neurons in each hidden layer, $d_\text{lattice}$ is the number of neurons in the output layer for predicting the deformation in the lattice, and $d_\text{coords}$ is the number of neurons in the output layer predicting the atomic displacements. A schematic of an MLP is shown in Figure \ref{fig:MLP_messagepassing}. The schematic of the whole graph neural network is shown in Figure \ref{fig:EGNN_schematic}}.
}
\label{tab:MLP}
\end{table}

\begin{figure}
\begin{center}

\begin{tikzpicture}[x=4cm, y=4cm, >=stealth]

\node[circle, draw, minimum size=1.2cm] (I1) at (0,.75) {$l^0_1$};
\node[circle, draw, minimum size=1.2cm] (I2) at (0,.25) {$l^0_2$};
\node at (0,0) {$\vdots$};
\node[circle, draw, minimum size=1.2cm] (I3) at (0,-.25) {$l^0_{d_{\text{i}}-1}$};
\node[circle, draw, minimum size=1.2cm] (I4) at (0,-.75) {$l^0_{d_{\text{i}}}$};

\node[] at (0,-1.25) {Input};
\node[] at (0,-1.35) {$\mathbf{l^0}$};

\node[circle, draw, minimum size=1.2cm] (H1) at (1,1) {$l^1_1$};
\node[circle, draw, minimum size=1.2cm] (H2) at (1,0.5) {$l^1_2$};
\node at (1,0) {$\vdots$};
\node[circle, draw, minimum size=1.2cm] (H3) at (1,-0.5) {$l^1_{d_{\text{h}}-1}$};
\node[circle, draw, minimum size=1.2cm] (H4) at (1,-1) {$l^1_{d_{\text{h}}}$};
\node[] at (1,-1.25) {Linear Layer 1};
\node[] at (1,-1.35) {$\mathbf{l^1} = \mathbf{W} \mathbf{l^0} + \mathbf{b} $};

\node[circle, draw, minimum size=1.2cm] (R1) at (2,1) {$l^2_1$};
\node[circle, draw, minimum size=1.2cm] (R2) at (2,0.5) {$l^2_2$};
\node at (2,0) {$\vdots$};
\node[circle, draw, minimum size=1.2cm] (R3) at (2,-0.5) {$l^2_{d_{\text{h}}-1}$};
\node[circle, draw, minimum size=1.2cm] (R4) at (2,-1) {$l^2_{d_{\text{h}}}$};
\node[] at (2,-1.25) {ReLU Layer};
\node[] at (2,-1.35) {$l^2_i = ReLU(l^1_i)$};

\node[circle, draw, minimum size=1.2cm] (L1) at (3,.75) {$l^3_1$};
\node[circle, draw, minimum size=1.2cm] (L2) at (3,0.25) {$l^3_2$};
\node at (3,0) {$\vdots$};
\node[circle, draw, minimum size=1.2cm] (L3) at (3,-0.25) {$l^2_{d_{\text{o}}-1}$};
\node[circle, draw, minimum size=1.2cm] (L4) at (3,-.75) {$l^3_{d_{\text{ot}}}$};
\node[] at (3,-1.25) {Linear Layer 2};
\node[] at (3,-1.35) {$\mathbf{l}^3 = \mathbf{W} \mathbf{r} + \mathbf{b} $};
\node[] at (3,-1.45) {$\mathbf{y} = \mathbf{l}^3$};

\node[circle, draw, minimum size=1.2cm, fill=gray!30] (O1) at (4,.75){$l^4_1$};
\node[circle, draw, minimum size=1.2cm, fill=gray!30] (O2) at (4,0.25){$l^4_2$};
\node at (4,0) {$\vdots$};
\node[circle, draw, minimum size=1.2cm, fill=gray!30] (O3) at (4,-0.25) {$l^4_{d_{\text{o}}-1}$};
\node[circle, draw, minimum size=1.2cm, fill=gray!30] (O4) at (4,-.75){$l^4_{d_{\text{o}}}$};
\node[] at (4,-1.25) {Optional ReLU Layer 2};
\node[] at (4,-1.35) {$l^4_i = ReLU(l^3_i))$};
\node[] at (4,-1.45) {$\mathbf{y'} = [l^4_1, l^4_2,...]$};


\foreach \i in {1,2,3,4} {
    \foreach \j in {1,2,3,4} {
        \draw[->] (I\i) -- (H\j);
        \draw[->] (R\i) -- (L\j);
        }
    }

\foreach \i in {1,2,3,4} {
    \draw[->] (H\i) -- (R\i);
    \draw[->, gray!50] (L\i) --(O\i);
}

\end{tikzpicture}
\caption{ Schematic of the message passing, vertex, lattice, and coordinate MLPs.  The input layer represented by $\mathbf{l^0}$ has {$d_{\text{input}}(d_{\text{i}}$)} dimensions, the hidden layers have {$d_{\text{hidden}}(d_{\text{h}}$)}  dimensions, and the output layer has {$d_{\text{output}}(d_{\text{o}}$)} dimensions. For message passing, lattice, and coordinate MLPs the MLP terminates in the second linear layer and the output is $\mathbf{y}$, {where this figure represents a single layer for the graph.} The last gray ReLU layer is only used in the vertex MLP where the output is $\mathbf{y'}$. For message passing and the vertex MLP {$d_\text{hidden}=d_\text{output}$}, for the lattice MLP {$d_\text{output}=9$} and for the coordinate MLP {$d_\text{output}=1$.}}
\label{fig:MLP_messagepassing}
    
\end{center}
\end{figure}

\begin{figure}[h!]
\centering
\begin{tikzpicture}[x=3cm, y=3cm, >=stealth]

\node[circle, draw, minimum size=1.2cm] (I1) at (0,1) {$\mathbf{h}^0_n$};
\node[circle, draw, minimum size=1.2cm] (I2) at (0,0.5) {$\mathbf{h}^0_{n-1}$};
\node at (0,0) {$\vdots$};
\node[circle, draw, minimum size=1.2cm] (I3) at (0,-.5) {$\mathbf{h}^0_2$};
\node[circle, draw, minimum size=1.2cm] (I4) at (0,-1) {$\mathbf{h}^0_1$};
\node[] at (0,-1.5) {Input Layer};

\node[circle, draw, minimum size=1.2cm] (H1) at (1.25,1) {$\mathbf{h}_n^1$};
\node[circle, draw, minimum size=1.2cm] (H2) at (1.25,0.5) {$\mathbf{h}_{n-1}^1$};
\node at (1.25,0) {$\vdots$};
\node[circle, draw, minimum size=1.2cm] (H3) at (1.25,-0.5) {$\mathbf{h}_2^1$};
\node[circle, draw, minimum size=1.2cm] (H4) at (1.25,-1) {$\mathbf{h}_1^1$};
\node[] at (1.25,-1.5) {Hidden Layer 1};


\node[circle, draw, minimum size=1.2cm] (H5) at (2.25,1) {$\mathbf{h}_n^L$};
\node[circle, draw, minimum size=1.2cm] (H6) at (2.25,0.5) {$\mathbf{h}_{n-1}^L$};
\node at (2.25,0) {$\vdots$};
\node[circle, draw, minimum size=1.2cm] (H7) at (2.25,-0.5) {$\mathbf{h}_2^L$};
\node[circle, draw, minimum size=1.2cm] (H8) at (2.25,-1) {$\mathbf{h}_1^L$};
\node[] at (2.25,-1.5) {Hidden Layer L};

\node[circle, draw, minimum size=1.2cm] (O0) at (3.25,0.5) {$\mathbf{h}_{out}$};

\node[circle, draw, minimum size=1.2cm] (O1) at (4,1) {$E_f$};
\node[circle, draw, minimum size=1.2cm] (O2) at (4,0) {$\mathbf{E}$};
\node[circle, draw, minimum size=1.2cm] (O3) at (4,-1) {$\mathbf{r_i^{\delta}}$};
\node[] at (4,-1.5) {Output Layer};


\fill[red, opacity=0.125]
plot[smooth cycle, tension=0.8] coordinates {
    (-0.1, 1.25) (-0.1, 0.25) (1.25, .75) (1.5, 1.25)
};
\fill[blue, opacity=0.125]
plot[smooth cycle, tension=0.8] coordinates {
    (1.35, -0.25) (1.35, -1.25) (0, -.75) (-0.1, -.25)
};

\foreach \i in {1,2} {
    \foreach \j in {1,2} {
        \draw[->] (I\i) -- (H\j);
    }
}
\foreach \i in {3,4} {
    \foreach \j in {3,4} {
        \draw[->] (I\i) -- (H\j);
    }
}
\foreach \i in {1,2} {
    \foreach \j in {5,6} {
        \draw[->] (H\i) -- (H\j);
    }
}
\foreach \i in {3,4} {
    \foreach \j in {7,8} {
        \draw[->] (H\i) -- (H\j);
    }
}
\foreach \i in {5,6,7,8} {
    \draw[->] (H\i) -- (O0);
    \draw[->] (H\i) -- (O3);
}
\draw[->] (O0) -- (O1);
\draw[->] (O0) -- (O2);

\end{tikzpicture}
\caption{Schematic of the graph neural network with $n$ input neurons, 2 hidden layers explicitly shown, ellipses indicating additional neurons, and 3 output neuron. {Only nodes within some vicinity are connected as represented by the graph.} Activation functions are applied at each hidden layer. Each hidden layer neuron is a vector of size $d_{\text{hidden}}$. {The vector is length $d_{\text{hidden}}$ because it comes from the output of the MLPs in Table \ref{tab:MLP} and Figure \ref{fig:MLP_messagepassing}}, $\mathbf{h}_n^L$ is the mean of all the output neurons which is used to calculate the global parameters: formation energy $E_f$ and the the strain $\mathbf{E}$. The set of atomic displacements $\mathbf{r}^\delta_i$ is a local parameter, and is size $n\times3$. Each arrow also encodes an MLP. {The} {red cloud shows} {which nearest neighbor nodes in the input layer contribute to node $\mathbf{h}_n^1$. Similarly, the {blue cloud shows} which nodes in hidden layer 1, the input node $\mathbf{h}_2^0$ contributes to. These show only {two} nearest neighbors for each node for simplicity, but this could be any number of nodes based on the radius that is set. }}
\label{fig:EGNN_schematic}
\end{figure}

\section{Results and Discussion} \label{sec:results}

\subsection{Background: DFT Computations and Trained Cluster Expansion}

In our previous work \cite{shojaei2024bridging}, we used the  Cluster's Approach to Statistical Mechanics software platform (CASM) \cite{Thomas2013Anharmonic,Puchala2013ZrO,VanderVen2018StatMech,casm} to choose 333 configurations and calculate their formation energies with DFT. {The formation energy was expressed as in Equation (\ref{eq:form_energy}) to vanish at $x=0$ and $x=1$. Figure 1c in \cite{shojaei2024bridging} reports the formation energies obtained from these DFT computations. Recall that the DFT convex hull is the envelope formed by the lowest formation energy, across all the configurations admissible at each composition} {up to 48 lattice sites}. The lowest formation energy across all configurations and compositions was obtained for a configuration with the zig-zag ordering, appearing as the minimum in the figure. A cluster expansion was then fit which had an MSE of 2.49 meV. All 333 configurations were used to train the cluster expansion {to ensure maximal predictive accuracy for the dataset. Using the complete set of configurations enables the cluster expansion model to learn from the full range of sampled configurations and capture all relevant interactions present in the DFT data.} 




\subsection{The EGNNs used}

\begin{table}[h!]
\centering
\begin{tabular}{|l|c|c|}
\hline
\textbf{Parameter} & \textbf{EGNN 1} & \textbf{EGNN 2} \\
\hline
Hidden Neurons ($d_\text{hidden}$) & 300 & 500 \\
\hline
Number of Hidden Layers & 5 & 5 \\
\hline
Learning Rate & 0.0001 & 0.0001 \\
\hline
Batch Size & 10 & 50 \\
\hline
Train/Test Split & 0.99 / 0.01 & 0.99 / 0.01 \\
\hline
Optimizer & Adam & Adam \\
\hline
Lattice Neurons ($d_{\text{lattice}}$) & N/A & 500  \\
\hline
Coordinate Neurons ($d_r$) & N/A & 500 \\
\hline
\end{tabular}
\caption{Comparison of EGNN training parameters for EGNN 1 predicting only formation energy and EGNN 2 predicting formation energy, strain, and coordinate displacements. }
\label{tab:EGNN_training_comparison}
\end{table}

We first created EGNN 1 to predict the formation energy, ignoring the lattice strain and atomic positions. This {represents a direct alternative} to the cluster expansion. We defined the graph based on the unrelaxed crystal structures and trained  EGNN 1 to predict the formation energy of the final (relaxed) configuration.

{The hyperparamters for training  EGNN 1 are shown in Table \ref{tab:EGNN_training_comparison}}.  A hyperparameter search indicated that this combination of parameters yielded the optimal testing and training MSE. Although a larger model could achieve a lower training MSE, it led to overfitting and worse performance on the testing set, suggesting that improved generalization would require additional data.

The results are shown in Figure \ref{fig: training only unrelaxed}. For EGNN 1, the RMSE (root mean squared error) for the testing data is $4.69$ meV and the RMSE for the training data is $0.18$ meV. In comparison, the cluster expansion training set had a RMSE of $2.49e$ meV. All configurations were used for training the cluster expansion, such that there was no testing set. The {order of magnitude}  lower training RMSE for  EGNN 1  compared to the cluster expansion and the comparable testing RMSE   compared to the training dataset for the cluster expansions suggests that  EGNN 1 is a viable method for replacing the cluster expansion and provides improved accuracy during training {while allowing lattice relaxation}. {Recall that there was no held-out testing data in Ref. \cite{shojaei2024bridging}.}

\begin{figure*}[t!]
    \centering
    \begin{subfigure}[b]{0.5\textwidth}
        \centering
        \includegraphics[width=\linewidth]{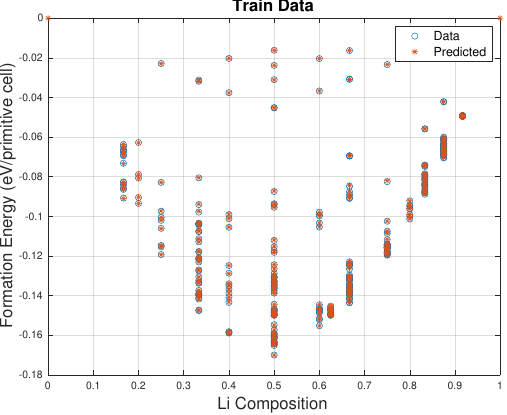}
        \caption{}
    \end{subfigure}%
    ~ 
    \begin{subfigure}[b]{0.5\textwidth}
        \centering
        \includegraphics[width=\linewidth]{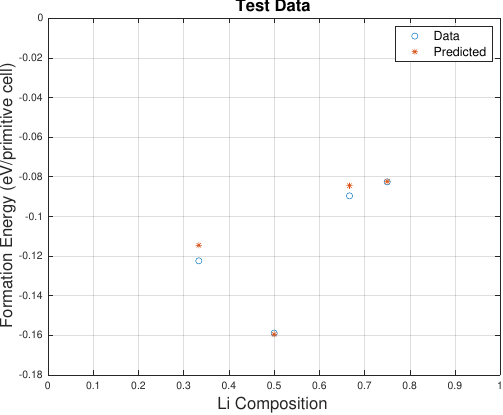}
        \caption{}
    \end{subfigure}
    \caption{{DFT data vs EGNN 1 formation energy predictions} for (a) the training set and (b) the testing set. }
    \label{fig: training only unrelaxed}
\end{figure*}

Next we set up  EGNN 2 to additionally predict the {lattice parameter deformation} {via the effective strain} and the {rotation-induced} change in atomic positions. The elastic constants for LCO were taken from Ref. \cite{sailuam2023influence}, at zero pressure. The elastic modulus values used are shown in Table \ref{tab:Cvalues}.
\begin{table}[]
    \centering
        \caption{The set of elastic constants (GPa) for  LCO where $C_{1144}$ was provided as -7 in the paper but we used a value of 0 since this coefficient corresponds to shear-normal coupling which we assume to vanish. For the rhombohedral symmetry of LCO, this elasticity tensor represents transverse isotropy in the 1-2 plane.\cite{sailuam2023influence}. }
    \label{tab:Cvalues}
    \begin{tabular}{|c|c|}
     \hline
        $C_{1111}$ &  344.71\\
        $C_{1122}$ & 109.14 \\
        $C_{1133}$ & 75.97 \\
        $C_{1144}$ & 0* \\
        $C_{3333}$ & 228.79 \\
        $C_{1313}$ & 50.73 \\
        $C_{1212}$ & 117.78\\
     \hline
    \end{tabular}

\end{table}

\begin{table}[]
    \centering
        \caption{RMSE Training and Testing results for  EGNN 2 for predicting the formation energy, the strain tensor (and corresponding strain energy) and the coordinate displacement. }
    \label{tab:RMSE}
    \begin{tabular}{|c|c|c|}
     \hline
        & Training RMSE &  Testing RMSE  \\
        \hline
        Formation Energy (eV/primitive cell) &  $9.83\times 10^{-4}$ & $4.38\times 10^{-3}$ \\
        Strain Energy (GPa/primitve cell) &  $6.86\times 10^{-6}$  & $2.02\times 10^{-4}$   \\
        Strain Tensor Values (GPa) &  $1.256\times 10^{-3}$  & $4.38\times 10^{-3}$  \\
        Coordinate Displacement (Bohr) &  $8.33\times 10^{-3}$  & $2.29\times 10^{-2}$ \\

     \hline
    \end{tabular}

\end{table}

\begin{figure*}[t!]
    \centering
    \begin{subfigure}[b]{0.5\textwidth}
        \centering
        \includegraphics[width=\linewidth]{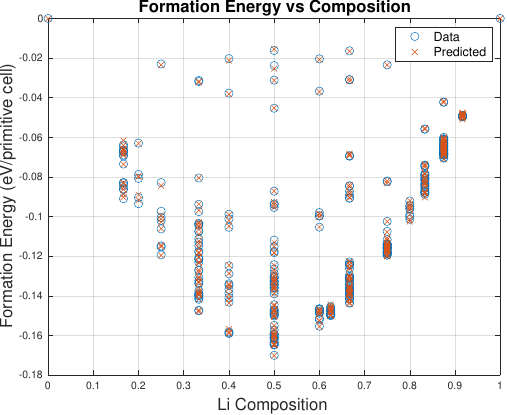}
        \caption{}
    \end{subfigure}%
    ~ 
    \begin{subfigure}[b]{0.5\textwidth}
        \centering
        \includegraphics[width=\linewidth]{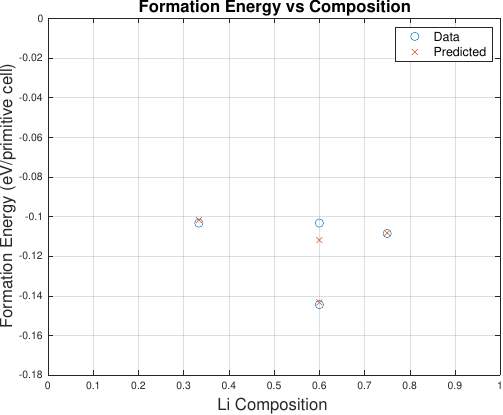}
        \caption{}
    \end{subfigure}
    \caption{DFT data vs EGNN predicted values for the formation energy for (a) training data (b) testing data. This figure shows results with  EGNN 2 which also predicts strain and coordinate displacement. }
    \label{fig: form_EGNN2}
\end{figure*}

\begin{figure*}[t!]
    \centering
    \begin{subfigure}[b]{0.5\textwidth}
        \centering
        \includegraphics[width=\linewidth]{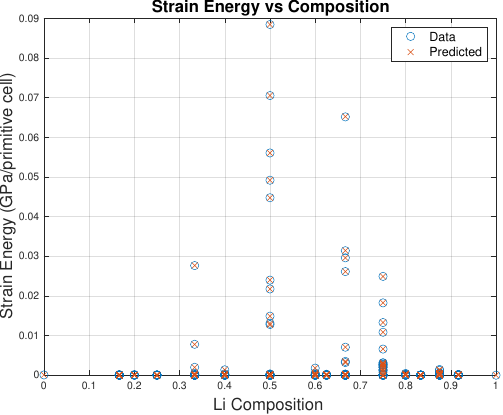}
        \caption{}
    \end{subfigure}%
    ~ 
    \begin{subfigure}[b]{0.5\textwidth}
        \centering
        \includegraphics[width=\linewidth]{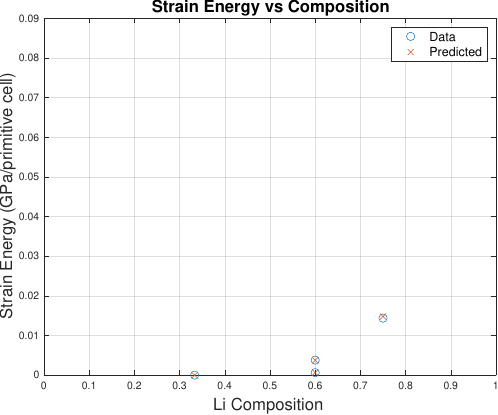}
        \caption{}
    \end{subfigure}
    \caption{DFT data vs EGNN 2-predicted values for the strain energy for (a) training data (b) testing data. }
    \label{fig: strain_EGNN2}
\end{figure*}

{The hyperaparemters for EGNN 2 are shown in Table \ref{tab:EGNN_training_comparison}. Additionally, the  output strain for EGNN 2 shown in Equation (\ref{eq:green-lagrange}) is multiplied by a factor of 100 to improve the training process. {This is an applied weight so the losses are of similar magnitudes to improve training.} 


The RMSEs {for the training and testing data as predicted by} the trained EGNN  are shown in Table \ref{tab:RMSE}. The formation energies vary between about $0$ and $ -0.17$ {eV} with an average of $\sim -0.1$ {eV} so both the training and testing RMSE are {reasonable} as shown in Figure \ref{fig: form_EGNN2}. The strain energies varied between 0 and $\sim 0.09$ {GPa} with an average of $\sim 6.6\times 10^{-3}$ {GPa}. { The RMSE of these strain energies is at least an order of magnitude below this average providing a good fit as seen in Figure \ref{fig: strain_EGNN2}. We also directly compared the strain tensor values. Given the maximum value of any component of the strain tensor from each member of the training (testing) set, we computed the average among all which is $1.53\times 10^{-2}$. Both the testing and training RMSE are at least an order of magnitude below this value, indicating the EGNN can provide a good approximation of the largest strain tensor values and correctly indicating if a strain tensor entry is small.} {The training and testing RMSE's are also shown in Table \ref{tab:RMSE}}

{The average coordinate displacement is $0.026$ Bohr, however this is not the most useful metric due to class imbalance and most displacements being close to zero. Instead, we can compare to the average maximum displacement of  $0.12$ Bohr and the maximum displacement overall of $0.37$ Bohr. With a testing RMSE of $2.29 \times 10^{-2}$ {Bohr}, we can expect to predict the largest displacements with a reasonable degree of accuracy and can predict when an atom will move a smaller amount.}{Coordinate displacement RMSEs appear in Table \ref{tab:RMSE}. }

\section{Conclusion}

This study demonstrates that equivariant graph neural networks (EGNNs) are  robust and flexible alternatives to traditional cluster expansions for predicting formation energies in crystalline solids. {Our results demonstrate that the EGNN provides a better representation of the formation energy, {with much more accurate results for training data predictions, and comparable predictions of testing data using the EGNN compared to training data for the cluster expansion.} and provides additional predictions including strain/atomic relaxations that traditional cluster expansions do not predict.}  By representing atomic configurations as graphs—where {vertices} correspond to atoms and edges represent interatomic bonds—EGNNs naturally capture the local chemical environments and structural motifs within a material. This graph-based representation also respects translational and rotational symmetries, which are fundamental to {lattice-based descriptions of materials.}

A key advantage of the EGNN approach is its ability to simultaneously predict multiple quantities of interest, including formation energy, effective strain over the supercell, and internal atomic displacements. These quantities are typically derived from computationally expensive DFT calculations. By learning to predict DFT outputs with a high level of accuracy, the EGNN can serve as a surrogate model that dramatically reduces the computational cost associated with large-scale simulations. For example, formation and strain energies can serve as inputs for Monte Carlo and phase-field simulations, while the strain tensor {captures effective lattice distortions over the computational cell used in the configuration}, and atomic displacements reflect relaxation behavior {internal to the configuration}. {The formation energy prediction exhibits slightly higher accuracy for  EGNN 1, which was trained exclusively on formation energy. This may be attributed to the increased complexity associated with simultaneously mapping multiple target properties with EGNN 2. Nevertheless, the second EGNN also demonstrates strong performance in predicting formation energies.} {To be explored is whether additional improvements in accuracy of EGNN 2 could be obtained by having more data for the multiple  properties targeted by it.}

Despite these strengths, several limitations remain. The model’s accuracy is constrained by the size and diversity of the training dataset. While strong performance was achieved for a specific subset of configurations, generalization to unseen compositions, defect-containing structures, or larger supercells may require more heterogeneous training data. Additionally, the current study does not address disordered or amorphous materials, where local environments vary significantly and pose greater challenges for graph construction and learning.

In conclusion, EGNNs provide a powerful framework for learning structure–property relationships directly from atomistic configurations. Compared to traditional cluster expansions, they offer greater flexibility, higher representational capacity, and the ability to model multiple outputs within a single architecture. Their incorporation of {periodic images} and their compatibility with diverse material systems make them well suited for future integration into multiscale modeling pipelines. The results presented here lay the groundwork for expanded applications, including defect prediction, structure relaxation, and high-throughput screening across chemically and structurally diverse materials.

\section*{Author Contributions}
J.H did most of the work of formulating, implementing and getting results with input by author's K.G. and S.S. All author's analyzed the results. J.H wrote the main manuscript text and all authors reviewed the manuscript.

\section*{Acknowledgments}
\label{sec:acknowledgements}
We gratefully acknowledge the support of Toyota Research Institute, Award \#849910, ``Computational framework for data-driven, predictive, multi-scale and multi-physics modeling of battery materials''.
This work used Expanse CPU at the San Diego Supercomputer Center (SDSC) through allocation MAT23005 and MCH240079 from the Advanced Cyberinfrastructure Coordination Ecosystem: Services \& Support (ACCESS) program, which is supported by U.S. National Science Foundation grants \# 2138259, \#2138286, \#2138307, \#2137603, and \#2138296.

\section*{Competing interests}
The authors declare no competing interests.

\section*{Data Availability}

The data that support the findings of this study are available from the corresponding author upon request.

\bibliographystyle{unsrt}
\bibliography{references}

\end{document}